\documentclass{article}
\usepackage[T1]{fontenc} 
\usepackage[utf8]{inputenc} 
\usepackage{ismir,amsmath,cite,url}
\usepackage{graphicx}
\usepackage{color}

\usepackage{lineno}

\title{EEG2Mel: Reconstructing Sound from Brain Responses to Music}

\twoauthors
  {Adolfo G. Ramirez-Aristizabal} {University of California Merced \\ {\tt aramirez64@ucmerced.edu}}
  {Chris Kello} {University of California Merced \\ {\tt ckello@ucmerced.edu}}

\begin{document}

\maketitle
\begin{abstract}
Information retrieval from brain responses to auditory and visual stimuli has shown success through classification of song names and image classes presented to participants while recording EEG signals. Information retrieval in the form of reconstructing auditory stimuli has also shown some success, but here we improve on previous methods by reconstructing music stimuli well enough to be perceived and identified independently. Furthermore, deep learning models were trained on time-aligned music stimuli spectrum for each corresponding one-second window of EEG recording, which greatly reduces feature extraction steps needed when compared to prior studies. The NMED-Tempo and NMED-Hindi datasets of participants passively listening to full length songs were used to train and validate Convolutional Neural Network (CNN) regressors. The efficacy of raw voltage versus power spectrum inputs and linear versus mel spectrogram outputs were tested, and all inputs and outputs were converted into 2D images. The quality of reconstructed spectrograms was assessed by training classifiers which showed 81\% accuracy for mel-spectrograms and 72\% for linear spectrograms (10\% chance accuracy). Lastly, reconstructions of auditory music stimuli were discriminated by listeners at an 85\% success rate (50\% chance) in a two-alternative match-to-sample task.     
\end{abstract}
\section{Introduction}\label{sec:introduction}
Reconstructing stimuli from brain responses has been explored across several related subdisciplines spanning from signal processing methodology, Cognitive Neuroscience, and deep learning modeling. A notable example comes from the culmination of experimental studies showing how to retrieve a Frequency Following Response (FFR) from averaged brainstem recordings to short ($<$ 1 second) presentations of speech and music \cite{skoe2010auditory,coffey2019evolving}. This type of recording only needs three electrodes, one centered on the scalp, a reference on the earlobe, and one grounding on the forehead. With this set up, participants are presented a sound such as a violin playing with their eyes closed repeatedly for about 1000 - 3000 times. The strength in this approach comes from the quality of FFRs to be able to contain the fundamental frequency, harmonics, and onset aligned amplitude from the auditory stimulus. This allows for researchers to playback an FFR and hear a lower resolution version of the original stimulus. Such a methodology provided an impressive proof-of-concept for speech and music retrieval, but it also depends on various experimental conditions. First, careful stimuli selection is necessary to retrieve FFRs as it needs to consider stimuli that have clear amplitude bursts, strong onsets, lower fundamental frequencies ($<$ 300 Hz), and short durations. Furthermore, the retrieved signal is averaged from hundreds repeated presentations used to isolate a specific response such as in brainstem activity. 

 Another exemplary approach in the space of signal processing and Cognitive Neuroscience comes from auditory tracking experiments where researchers have been successful in retrieving the speech amplitude from the stimuli \cite{synigal2020including}. Experiments had participants listening to 180 secs of an audiobook split up into 25 separate trials. What makes this approach strong is the simplicity behind the modeling procedure where it takes as input both low frequency and high-gamma power spectrum signals from Electroencephalogram (EEG) channels and puts them through a linear model that decodes the input and then maps to the stimulus target. Despite the modeling simplicity, the approach faces a trade off with the data processing needing to be more complicated. EEG signals need to first be split into two signals that separate activity from low and high frequencies while also being iteratively integrated across channels into the model over several time lags from 0-250 ms. Nevertheless, this research moves forward the methodology of neural decoding and acoustic information retrieval from cortical activity towards a more naturalistic scenario where it does not require mass averaging of presentations and stimuli presentations are longer. 

 \begin{figure*}[t]
 \centering
\includegraphics[width = 16cm, height = 4.4cm]{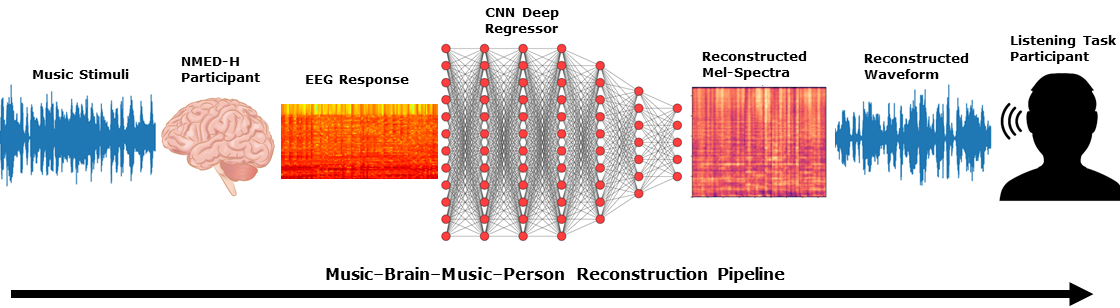} \\
\label{f:name}
\caption{\small Visualization of music reconstruction in our study. Brain responses from music listening are processed by deep regressors and retrieved music is played back to new participants.}
\end{figure*}
Deep learning methods of decoding stimuli from brain responses have been successfully implemented through various classification models. In the image stimuli domain, researchers collected passive cortical responses of people watching images from ImageNet \cite{spampinato2017deep}. They trained the EEG responses in a Long Short-Term Memory (LSTM) based Recurrent Neural Network (RNN). Model validation demonstrated that brain responses could be used to correctly identify an image class at 84\% performance that is being perceived (2.5\% chance), i.e., if a participant was watching an image of a panda the model could correctly identify that the image being perceived was of a panda. In the auditory domain, recent classification studies have developed methods to efficiently use EEG responses in Convolutional Neural Networks (CNNs) to correctly classify the name of the song a participant is listening to. This was shown to work strongly (85\%) when participants listened to music they were familiar with (8.33\% chance), and if the input of the models were formatted as 2D representations of the power spectrum density across channels \cite{sonawane2021guessthemusic}. Using different datasets, another study replicated the prior study’s results and further evaluated the efficacy of EEG data being used as images in computer vision models implemented through a custom version of AlexNet architecture and in transfer learning with a pretrained ResNet model. The researchers trained models with the power spectrum density EEG representations and compared it to the raw EEG voltage representations. Training on the raw input representation was able to show that not only was end-to-end classification possible, but that it could also achieve State-of-the-Art performance at ~88.6\% with 10\% chance on unfamiliar music \cite{ramirez2022image}.

\begin{table}[h]
    \centering
    \scalebox{0.8}{
    \begin{tabular}{c|c c c}

    \hline
    Layer Type & Filter Size & Input & Output\\
    \hline
    \hline
    Conv2D & $4 \times 4$ & $63 \times 125 \times 1$ & $63 \times 125 \times 8$ \\
    BatchNorm2D & - & - & - \\
    \hline
    Conv2D & $4 \times 4$ & $63 \times 125 \times 8$ & $63 \times 125 \times 16$ \\
    BatchNorm2D & - & - & - \\
    \hline
    Conv2D & $4 \times 4$ & $63 \times 125 \times 16$ & $63 \times 125 \times 32$ \\  
    BatchNorm2D & - & - & - \\
    \hline
    Conv2D & $4 \times 4$ & $63 \times 125 \times 32$ & $63 \times 125 \times 64$ \\  
    BatchNorm2D & - & - & - \\
    \hline
    Conv2D & $4 \times 4$ & $63 \times 125 \times 64$ & $32 \times 63 \times 128$ \\  
    BatchNorm2D & - & - & - \\
    \hline
    MaxPool2D & $2 \times 2$ & $32 \times 63 \times 128$ & $17 \times 32 \times 128$ \\
    Flatten & - & $17 \times 32 \times 128$ & 69632\\
    \hline
    FC1 & - & 69632 & 128 \\
    BatchNorm1D & - & - & - \\
    FC2 & - & 128 & 5632 \\
    \hline
    Reshape & - & 5632 & $44 \times 128$\\
    \hline
    \end{tabular}}
    \caption{\small Architecture used in our deep regressors. This specific model was trained on the NMED-H dataset with a spectral input and mel-spectra music target. }
    \label{t:architecture}
\end{table}
 \begin{figure*}[t]
 \centering
\includegraphics[width = 10.55cm, height = 5.55cm]{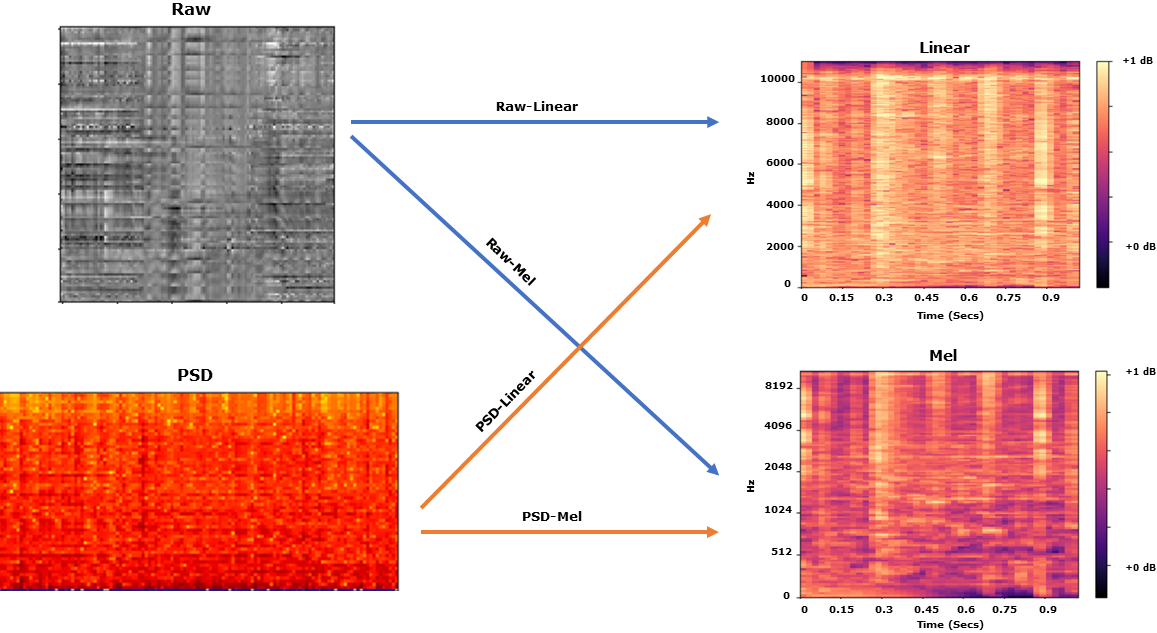} \\
\label{f:name}
\caption{\small On the left are the two types of input representations we test,  and on the right the two types of target representations for a total of 4 model combinations as labeled by each arrow. All representations come from Participant 1 at the 100th second.}
\end{figure*}
Studies moving beyond classification approaches and into trying to reconstruct the stimuli have seen some initial success with the use of generative models and careful feature extraction. Following the work in \cite{kavasidis2017brain2image}, the researchers transferred weights from their classification models to use as encoders in a Variational Autoencoder (VAE) and Generative Adversarial Network (GAN). They successfully recreated the image classes that participants were looking at with an EEG response as input to the network. For example, if a participant was looking at images of pandas, then the models would recognize that it is a panda and output an image of a panda. The novelty of generative models comes from the ability to not just have outputs that are semantically similar, as their study claims, but to also generate a distribution of images not directly presented to participants. This was made possible by the transfer learning of weights trained to classify image classes, which when fine-tuned to new examples created training procedures of generative models estimating the density of image classes. For the purposes of direct information retrieval this also becomes a limitation because the outputs are estimations from image classes rather than a pixel-by-pixel reconstruction. This means that if a participant was looking at a picture of a panda who was sitting down and profiled to the left, then the model would not necessarily pick up on those features but return a sampled panda image from the estimated image class fitted by the model. In music retrieval, another study had found a way to leverage the acoustic features of the music stimuli and use it as inputs for a multi-view approach in a deep Variational Canonical Correlation Analysis (VCCA) model \cite{ofner2018shared}. Despite the lack of quantitative model validation analyses, this study demonstrated the possibility of using deep learning to be able to reconstruct music spectra from long EEG responses. In their qualitative analyses, they confirmed that reconstructed spectra were able to retain acoustic features from the original stimuli such as pitch, timbre, and tempo. Furthermore, it allowed for an approach that would move beyond stimuli classes and attempt to consider time-aligned activity in the stimulus presentation to be retrieved from brain responses. Here, we present results that advance the methodology of music retrieval and stimuli reconstruction from brain responses by training models that have time aligned EEG responses to the music spectra target. This is done without needing to integrate acoustic features from the stimuli as input to the model or depending on a multi step feature extraction process from the EEG. We also present a series of quantitative validation methods to measure the success of music reconstruction, including feedback from participants listening to the reconstructed music from our model outputs. Figure 1 outlines the scope of information retrieval, starting with music stimuli presented to participants being encoded in their EEG responses and mapped to the time-aligned stimuli spectrogram by a deep regressor which outputs a reconstructed music spectrogram that is inverted into a waveform and presented to a new person.  

\section{Methods}
\subsection{Datasets}\label{subsec:datasets}
Here we train and validate models with the Naturalistic Music Electroencephalogram Dataset – Tempo (NMED-T) and Naturalistic Music Electroencephalogram Dataset – Hindi (NMED-H) \cite{losorelli2017nmed,kaneshiro2016naturalistic}. Both datasets collect long recordings of passive brain responses to participants listening to music. The NMED-T contains recordings from twenty participants listening to ten different songs that were selected by the researchers because of the lack of familiarity and differences in tempo. On the other hand, the NMED-H has recordings from twelve participants listening to four pop songs in Hindi. These publicly available datasets have been central to various signal processing studies along with some studies looking into music retrieval from cortical activity \cite{ofner2018shared,vinay2021mind}. To aid with replication we provide guiding python notebooks as examples to our methods and analyses for others to follow. The supplementary materials also provide examples for others to listen to and visually inspect. Both datasets provide preprocessed versions of the data which include standard corrections for faulty channels, line-noise filtering, and muscle artefact corrections \cite{losorelli2017nmed,kaneshiro2016naturalistic}. Here we use those preprocessed versions of the data, as our focus is on passive cortical signals and not on any correlated motor behaviors that may show up as muscle artefacts. 

\subsection{Model \& Training}\label{subsec:model}
Our modeling approach contrasts with previous publications showing success with decoding and reconstructing complex stimuli, where generative models such as VAEs and GANs reconstructed image classes and where VCCA was able to reconstruct music stimuli spectra at the expense of needing a multi-view of extracted features \cite{kavasidis2017brain2image,ofner2018shared}. We choose a straightforward approach with a sequential CNN based regressor mapping EEG input directly to the time-aligned music spectra. Table 1 shows a summary of how the model architecture is constructed; the models contain five convolutional layers with the last convolutional layer being the first layer that reduces the dimensionality of the input. A Max Pooling layer with a small pool size is chosen over a Global Average Pooling layer used previously during EEG classification \cite{ramirez2022image} using the same dataset, because it was necessary to limit shrinkage since the output layer was the size of the spectral target. 

Regularization in the model was implemented through dropout layers and maintaining the size of intermediate fully connected layer as small as possible. Dropout layers between convolutional layers were kept with 10\% dropout and the dropout layer between the fully connected layers was kept at 15\%. The amount and strength of dropout layers were tested demonstrating no evident effect with less layers and weaker layers, stronger values were not helpful between convolutional layers but did show some improvement between fully connected layers. Activation and Kernel L2 regularization was tried but opted out due to showing signs of a vanishing gradient. Number of filters for convolutional layers were kept at a base 8 while increasing by a power of 2 for every subsequent layer. Increasing the number of filters showed training loss outpacing validation loss resulting in overfitted model runs, while decreasing the number of filters made training loss stagnate too early which resulted in underfitting. For all intermediate layers a Rectified Linear-Unit activation was used with a linear output activation. Non-monotonic activation functions Swish and Mish were tried due to their success in improving image processing in deep networks, but they did not present any evident advantage over ReLu during training. Lastly, we use Adaptive Moment Estimation (Adam) as an optimizer with a 0.0015 learning rate and initialized weights with a He uniform distribution which have shown advantages in similar training procedures \cite{ramirez2022image,ebrahimpour2020end}.

The first four minutes of all recordings were used and cut up into five second chunks. To balance the train and test set distributions across time, we assign every other chunk to either train or test at a 75/25 ratio. Then all chunks were split into 1 second examples and randomly shuffled for training and validation. Five second chunks were also useful in securing consecutive 1 second examples to be reconstructed. These reconstructed five second music spectra were inverted into waveforms and used as examples in a behavioral experiment to validate the quality of brain to music reconstruction. Generalization stays within participants and across unseen chunks of time that are balanced to be sampled from the beginning, middle, and end of the songs. Other EEG studies also keep generalization within participants \cite{moinnereau2018classification,sonawane2021guessthemusic}, and a recent study has shown that weak correlations across participants in the NMED-T could be why generalizing to unseen participants is difficult without many recorded participants in the dataset \cite{pandey2022music}. 

\section{Experimental Results}\label{sec:results}
\subsection{Representations}\label{subsec:reps}
Using the NMED-T, four reconstruction models were trained and evaluated depending on their varying input and target representations. Figure 2 outlines the combinations between input and target representations along with their visual qualities. Each arrow in the figure is a model showing the designated input to target mapping. Prior classification studies have shown a preference for input representations either being raw voltage or a Power Spectral Density (PSD), and their ability to boost performance in CNNs \cite{sonawane2021guessthemusic,ramirez2022image}. As a regression task, we also find it important to compare target representations since linear and mel-spectrogram representations have tradeoffs. The linear spectrum scales low frequency activity equally to higher frequency, which allows it to be more easily inverted back into a listenable waveform. On the other hand, the mel-spectrogram applies a non-linear scaling across frequency ranges that has made it easier to map in classification tasks \cite{gururani2018instrument,chillara2019music}, but also adds more noise during the inversion to a listenable waveform. 
\begin{table}[t]
    \centering
\scalebox{1.05}{
\begin{tabular}{c|c|c}
\hline
 \textbf{Inputs (1 secs)}& \textbf{Targets (1 secs)}& \textbf{Accuracy}\\
 \hline
 PSD (63,125) & Mel-Spec (44,128) & 80.80\%\\
 \hline
 PSD (63,125) & Lin-Spec (44,1025) & 72.28\%\\
 \hline
 Raw (125,125) & Mel-Spec (44,128) & 46.07\%\\
 \hline
 Raw (125,125) & Lin-Spec (44,1025) & 37.53\%\\

 \hline
\end{tabular}}
    \label{t:ours_diff_ds}
\caption{\small Summary of reconstruction models' output classification across the four representation combinations. Each representation is shown with its data shape for 1 sec in parentheses. }
\end{table}
The reconstruction models were all trained until they reached a plateau in their loss with varying epoch ranges (~100-300) needed for the models to converge their training and validation performance. Given that the mean-squared error (MSE) loss was only indicative of the convergence of the model during training and not a metric that informs how image reconstruction compares across models, we decided to classify the output as an objective quantitative measure. Simply put, the four trained regressors were used to create new datasets from their outputs given the original NMED-T as input and keeping the same train-test split. Then those new output datasets were passed through a classifier that would attempt to classify the name of the song from each spectral image reconstruction. This type of validation has been used before in the image stimuli domain when trying to objectively test the quality of reconstructed image classes from generative models
\cite{kavasidis2017brain2image}. The classifiers were CNN based following the architecture from a previous study where the EEG was classified into song name classes \cite{ramirez2022image}. Table 2 demonstrates a summary of the results from the classified outputs across modeling approaches. The raw input representation showed to not give the best results in our deep CNN regressors but it did remain well above the chance performance rate of 10\%. Meanwhile, the mel-spectrogram worked better across both the raw and PSD input representations with performance comparable to studies where only EEG classification was conducted \cite{stober2014using,moinnereau2018classification,yu2018deep}. 
\begin{figure}[t]
\centering
\includegraphics[width = 6cm, height = 5cm]{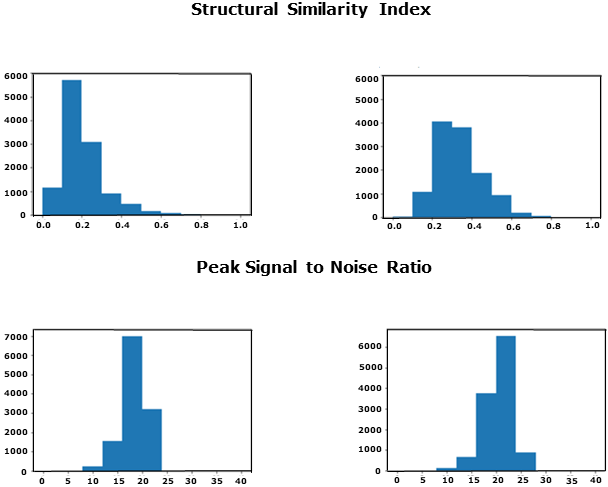} \\
\label{f:name2}
\caption{\small Distributions of SSI and PSNR scores across target representations, along with their mean score.}
\end{figure}

 \begin{figure*}[t]
 \centering
\includegraphics[width = 10.55cm, height = 5.55cm]{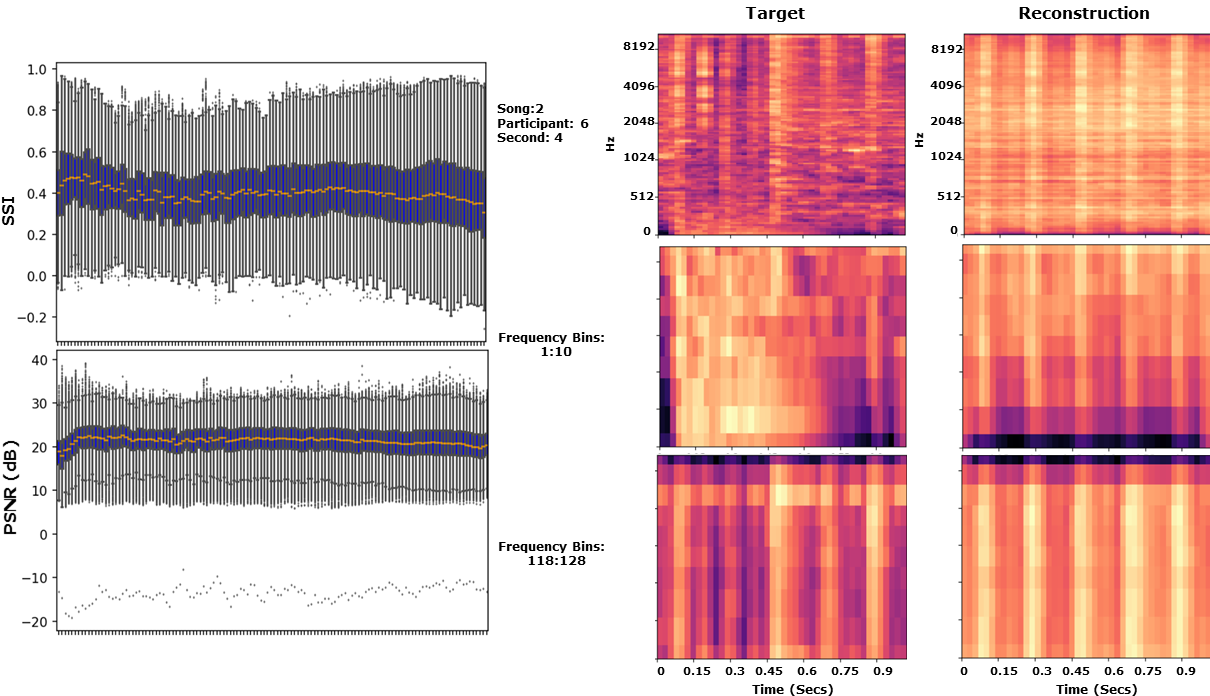} \\
\label{f:freq}
 \caption{\small SSI and PSNR scores across the 128 mel frequency bins of all test examples. On the right is a reconstructions example zooming in to the lower and higher frequencies. }
\end{figure*}

Classifying outputs to their stimuli classes reveals fidelity of semantically relevant reconstructions. This was especially useful in the image stimuli domain when the models being evaluated focused at the stimuli class level \cite{kavasidis2017brain2image}. Furthermore, their results of classification stayed well above chance (2.5\% chance) for forty classes with performance \~40\%. The results from Table 2 also stand well above chance for 10 classes (10\% chance). Given that our training method maps EEG to the target stimuli directly rather than estimating stimuli class densities in the model's latent space, common image reconstruction metrics were used to see how different each reconstruction was from the target image. Structural Similarity Index (SSI) and Peak Signal to Noise Ratio (PSNR) were chosen because of the ability to compare the reconstruction of perceived image features and how much noise the reconstructions contain from a lossy compression comparison respectively \cite{hore2010image}. Figure 3 summarizes the results from both metrics across the test set of the image reconstructions from the regressor models trained on the PSD input representation along with their linear and mel-spectrogram targets. SSI measures, as a normalized ratio, the similarity between images where an SSI score of 1 points to the images being identical. The mean SSI is higher in the mel-spectrogram reconstruction than with the linear-spectrogram by ~14\%. PSNR on the other hand, measures as a ratio the logged maximum image power over its mean squared error which gives a relative difference between images and not a normalized score. We also see in Figure 3 that mel-spectrogram has a higher mean PSNR which further validates how the mel-spectrogram makes the EEG to music spectra reconstruction closer to the target.  
 
To better understand what musical features the model learned from mapping to the mel-spectrogram, we took SSI and PSNR scores across frequencies. Figure 4 demonstrates the results of this analysis; for every example we paired up each 1 second frequency bin from target to model output to calculate both an SSI and PSNR value. Boxplots were created for all 128 frequency bins, and Figure 4 shows that frequencies vary around a 0.4 SSI score. Meanwhile, the PSNR plot shows notably lower values for the first 5 frequency bins. We show a visual example of this difference by taking a reconstruction and zooming into the ten lowest and highest frequency bins. This visual comparison demonstrates how many more pixel differences the lower frequencies have when compared to the higher frequencies. This went against our expectations as we believed it would be easier to map to lower frequencies because of pattern regularity \cite{lenc2018neural}. Further testing is needed as future studies adopt this methodology but we speculate that this could be attributed to these model architectures not explicitly learning temporal dependencies such as in an RNN.  
\subsection{Spectra to Music}\label{subsec:spec}
The above-mentioned metrics were useful in evaluating the success of the spectral image reconstructions, but here we are also interested in whether that allows for the retrieval of perceptually interpretable music reconstruction. This means that our spectral image reconstructions should be good enough to then be processed by common out of the box signal processing libraries that invert spectrograms to soundwaves, and then be identified by listeners as matching the original stimulus. For a practical experimental design and for testing the generalization of our methodology, we decided to train models and produce reconstructed spectral outputs from the NMED-H. Instead of a total of ten songs like in the NMED-T, the four songs from the NMED-H made it easier to test reconstructions across time slices and recorded participants; the training procedure in the deep CNN regressors stayed the same. Models trained on this dataset converged sooner and provided a steeper gradient descent during training which could be attributed to the familiarity of songs from participants \cite{hadjidimitriou2013eeg}. Because the best performing models using the NMED-T used the PSD and mel-spectrogram as input-target representations, we focused our training on the NMED-H with that specific representation pairing. Figure 5 provides a visual summary of the quality of reconstructions from the NMED-H.
\begin{figure*}[t]
 \centering
\includegraphics[width = 10.55cm, height = 5.55cm]{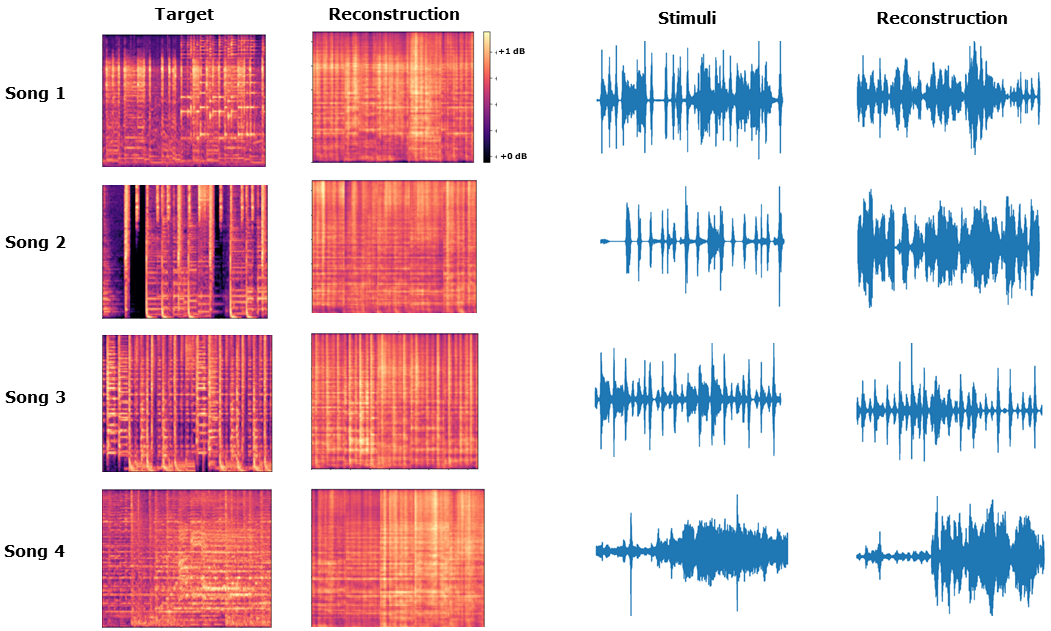} \\
\label{f:rec}
 \caption{\small Five second examples of model spectra predictions (left) and their reconstructions from spectra to sound wave (right). Examples come from the 10th second across randomly selected participants. }
\end{figure*}

 As a tradeoff for efficient modeling procedures and feature mapping within the proposed methodology, processing the outputs to listenable waveforms faces several steps of lossy transformations. The first comes from the actual deep learning models themselves where the input to target mapping assumes that the music signal is hidden/scrambled in cortical activity and attempts to recover that into a spectral representation. Second, those model outputs must be denormalized and set to a decibel range. Lastly, the spectrograms are transformed into waveforms by approximating the Short-Time Fourier Transform (STFT) magnitude from the mel power spectrum and reconstruction of the phase is done using Griffin Lim Algorithm (GLA). The mel-spectrogram in comparison to the linear is more lossy in the last two steps because mel is scaled as a non-linear function that represents human auditory perception across frequencies. Nevertheless, our modeling procedure finds it easier to map mel targets, so we commit to lossy tradeoffs and adjust to produce listenable examples to present to participants during the behavioral evaluation. During the denormalization step we use the following transformation:
 \begin{equation}
\operatorname{Mel\,dB} = (\operatorname{mel}^T \times \operatorname{Max\,dB}) - \operatorname{Max\,dB} + \operatorname{ref\,dB}
\end{equation}

In Equation 1, $\operatorname{mel}^T$ is the mel-spectrogram we aim to denormalize set as a matrix transpose and $\operatorname{Max\,dB}$ is set as a constant to 100 while $\operatorname{ref\,dB}$ is set as a constant within a song but varies across songs. To avoid presenting jarring noise to participants, $\operatorname{ref\,dB}$ becomes a free parameter set to [46 dB, 46 dB, 52 dB, 43 dB] for songs 1-4 respectively. The parameter value was set simply by randomly sampling 5 second examples and listening to how they would sound with values between 40 dB – 60 dB. The evaluation of which value produced the least jarring example was decided by the researcher; therefore we encourage any other studies following this methodology to play around with their own constraints relevant to their own studies. In Figure 5 it is demonstrated how well the waveforms come out despite the lossyness of the inversion process. The reconstructed spectrograms in Figure 5 shows how information across frequencies are recovered and how that translates to alignment in amplitude from target to reconstructed waveform.

The produced listenable examples were then presented to participants in a listening task to evaluate whether their quality was good enough to be identifiable. The experiment was a two-alternative AB-X task in which a participant had to pick between sounds A or B that matches the unlabeled sound X. The target sound X was a 5 second example from one of the four song stimuli in the NMED-H. Sounds A or B contained a non-corresponding foil and a corresponding 5 second reconstruction from our deep regressor model that was taken from spectra to a waveform. Participants were tested on a total of 24 trials plus 4 practice trials at the start of the experiment. The reconstructed model output produced a total of forty-eight examples which then half of the examples were used for one version of the experiment and the other half for a second version. Each version of the experiment controlled for balancing presentations across time by picking examples that belonged to the beginning, middle, and end of the songs. Furthermore, the experiment also contained the same number of examples per song while controlling for each trial to have reconstructions coming from different participants from the NMED-H. Trials were all randomized and foils balanced to minimize the repetition of reconstructions from a specific song or participant. During the AB-X discrimination task, participants were given a max of 30 minutes to finish and were allowed to listen to the sound samples as many times as they liked before making their choices, while also not taking longer than 1 minute per presentation. This experiment collected responses from a total of 16 participants with 8 participants for each version of the task. On average, participants had an 85\% success rate (50\% chance) with max performance of 95.83\% and minimum performance of 66.67\%. The average performance lines up with output classification, adding evidence of not just robust image reconstruction but also the perceived interpretability of listeners when inverting the spectrum.
\section{Conclusion}\label{sec:conclusion}
Here we show that it is possible to reconstruct music presented to participants using their EEG responses as input to a quality that allows others to correctly identify it when they listen to the music recovered from a brain response. The EEG to music stimuli mapping approach allows for reconstructions to preserve time-dependencies, during concatenated examples, necessary in the perception of music rather than taking an average of the stimuli class. Furthermore, no heavy feature extraction from the EEG or stimuli signals are needed in this process and only requires the EEG input to be transformed into a power spectrum for high quality reconstructions. With this we have shown that a computer vision approach has allowed for the processing of EEG responses to long naturalistic music stimuli as well as the ability to recover information across all frequency components. Finally, our results present a unique case for computer vision tasks, where what are normally treated as time-series (EEG and Music) are represented as images for more efficient training procedures. This leaves a path open for future investigations focusing on the efficacy of specific computer vision architectures being tasked to retrieve acoustic information from noisy signals like EEG.
\\
\\
\\
\\
\\

\bibliography{ISMIRtemplate}

%
%
%
%
%

\end{document}